\documentclass[a4paper,11pt]{article}
\usepackage{amsmath,epsfig,graphicx}
\newcommand{\grad}{{\rm grad}}

\begin{document}

\begin{center}
{\Large\bf Kelvin force in a Layer of Magnetic Fluid}\\[.5cm]
Adrian Lange\\[.5cm]
Institut f\"ur Theoretische Physik, Universit\"at Magdeburg, Postfach 4120, D-39016 Magdeburg, Germany\\
\end{center}

\vspace{.2cm}

\begin{abstract}
The Kelvin force in a layer of magnetic fluid subjected to a homogeneous magnetic field and local
heating is studied. The study is motivated by the question about the corresponding Kelvin force
density [M. Liu, Phys. Rev. Lett. {\bf 84},  2762 (2000)]. It is shown that the usual and the 
newly proposed formulation of the Kelvin force are entirely equivalent. It is only when
approximations are introduced that differences arise.
\end{abstract}

{\bf Keywords:} Kelvin force; Magnetic fluids;\\

\begin{tabular}{ll}
{\bf Contact author:}& Adrian Lange, Fax:  +49-391-6711205\\
 & \makeatletter email: adrian.lange@physik.uni-magdeburg.de \makeatother
\end{tabular}

\newpage

Two prominent directions of interest can be identified among the present studies of
phenomena occuring in magnetic fluids (MFs). These are the study of
unconventional fluid dynamics phenomena such as the ``negative viscosity'' effect and
the Weissenberg effect (for a review see \cite{odenbach00}) and the proposal of
new theoretical concepts \cite{liu_HMT}. Both directions are interwoven in the observation
of a novel convective instability in a horizontal MF layer \cite{du98,luo99} and in
the subsequent discussion about the correct form of the magnetic (or Kelvin) force density
\cite{liu_comment00,luo_comment00}.

This first discussion about the range of validity of the Kelvin force was followed by
a second one \cite{engel01_comment,liu01_reply} triggered by a paper announcing that
a pendulum experiment had confirmed the invalidation of the Kelvin force in MFs \cite{odenbach01}.
The claim was not accomplished
according to \cite{engel01_comment} and the resulting need for a clarification entailed
an extended paper \cite{engel01}. A clarification as in the pendulum experiment 
is lacking for the convection experiment. The aim of the present paper is to show that
the usual and the proposed formulation of the Kelvin force are entirely equivalent.
It is only when one introduces approximations that differences arise.
 
In \cite{du98,luo99} a horizontal layer of MF (stable colloidal suspension of
magnetite nanoparticles dispersed in kerosene) between two glass plates is locally
heated by a focused laser beam. It passes perpendicularly through
the layer in the presence of a homogeneous vertical magnetic
field. The absorption of the light by the fluid generates a temperature
gradient and subsequently a refractive index gradient. This gradient is
optically equivalent to a diverging lens, leading to an enhancement of the
beam divergence. As result, depending on the strength of the magnetic field
different diffraction patterns appear \cite{luo99}.

To explain the observed phenomena, the form of the magnetic force inside the
fluid has to be known. Therefore a horizontal layer of MF is considered which is
subjected to a homogeneous vertical magnetic induction. Since the temperature and
the concentration of the fluid may vary, the magnetic field in the fluid is
inhomogeneous and give rise to a finite Kelvin force density, ${\bf f}_{{\rm K}}$.
It can be derived from the Helmholtz force \cite{landau}
\begin{equation}
\label{eq:1}
{\bf f}_{{\rm K}} =\mu_0\grad\left[ {H_{int}^2\over 2}\rho{\partial\chi\over
\partial\rho}\right]-\mu_0{H_{int}^2\over 2}\grad\chi \, ,
\end{equation}
where $\chi=\alpha\rho (1+\beta_1\alpha\rho)$ is the susceptibility of the MF,
$\rho$ its density, $H_{int}$ the absolute value of the magnetic field inside the fluid,
and $\mu_0=4\pi\times 10^{-7}$ H/m. Higher order terms in $\rho$ are included in $\chi$
in order to determine the Kelvin force beyond the dilute limit. This limit is given by
$\beta_1\equiv 0$, i.e. $\chi =\chi_L$ which is the susceptibility according to Langevins
theory which assumes non-interacting particles. In this approximation $\chi_L$ depends
linearly on the density ,
$\chi_L = \alpha\rho =\mu_0 m^2 \rho/(3 k T m_{eff})$, where $m_{eff}$ is the effective
mass of a ferromagnetic particle with its `attached' carrier liquid molecules
\cite{bashtovoy88}, $m$ the
magnitude of the magnetic moment of the particles, $T$ the temperature, and $k$ the
Boltzmann constant. The coefficient $\beta_1$ of the quadratic term in $\rho$ was determined
in different microscopic models \cite{onsager36,mean_spheric,buyevich92} which all
provide the same value $\beta_1=1/3$. In the presence of a uniform external magnetic
induction ${\bf B}_{ext}$, the internal field is given by
${\bf H}_{int}={\bf B}_{ext}/(\mu_0(1+\chi))$. Inserting all expressions in
Eq.~(\ref{eq:1}), the Kelvin force follows as
\begin{equation}
\label{eq:2}
{\bf f}_{{\rm K}}(\chi) =-{B_{ext}^2\over \mu_0}{\chi_L^2\left\{1+\beta_1\left[3\chi_L\left( 1+
\beta_1\chi_L\right)-1\right]\right\}\over (1+\chi)^3}{\grad\chi_L\over \chi_L}\, .
\end{equation}

In \cite{liu_comment00} a variant form for the Kelvin force is proposed.
By defining a different
susceptibility $\bar\chi$ via ${\bf M}=(\bar\chi/\mu_0){\bf B}_{int}$ with
$\bar\chi=\chi/ (1+\chi)$
the Helmholtz force has now the variant form \cite{liu_comment00}
\begin{eqnarray}
\nonumber
{\bf f}_{{\rm V}} &=&\grad\left[ {B_{int}^2\over 2\mu_0}\rho{\partial\bar\chi\over
\partial\rho}\right]-{B_{int}^2\over 2\mu_0}\grad\bar\chi \\
\label{eq:4}
&=&{B_{int}^2\over 2\mu_0}\grad\left[ \rho{\partial\bar\chi\over
\partial\rho} -\bar\chi\right]+{\rho\over 2\mu_0}{\partial\bar\chi\over
\partial\rho}\grad B_{int}^2 \, .
\end{eqnarray}
${\bf B}_{int}$ (${\bf M}$) is the magnetic induction (magnetization) in the fluid
and $B_{int}$ its absolute value. Due to the uniform form of the external induction
and the continuity of the magnetic induction across the interface, the last term is
zero and the first term gives with the definition of $\bar\chi$
\begin{equation}
\label{eq:5}
{\bf f}_{{\rm V}}(\chi) ={B_{ext}^2\over 2\mu_0}\grad\left[ {\beta_1\alpha^2\rho^2 -\chi^2
\over (1+\chi)^2}\right]\, .
\end{equation}
Executing the differentiation in Eq.~(\ref{eq:5}) leads exactly to the same result
as in Eq.~(\ref{eq:2}). Therefore both formulations are indeed physically equivalent
under the inclusion of a quadratic term in $\rho$. The equivalence is true also for
higher terms in $\rho$ provided that the susceptibility can be written as
$\chi=\alpha\rho\left[ 1+\sum_{i=1}^\infty\beta_i (\alpha\rho)^i\right]$. Thus there
is no a priori reason to prefer Eq.~(\ref{eq:4}) over Eq.~(\ref{eq:1}) because
both formulations lead to the same result as long as the definition of $\bar\chi$ is
used in Eq.~(\ref{eq:4}). This very basic fact independent on the
relation of $\chi$ on $\rho$ has to be emphasized since it recedes in the wake of
the discussion \cite{liu_comment00,luo_comment00}.

\begin{figure}[htbp]
  \begin{center}
    \includegraphics[scale=.5]{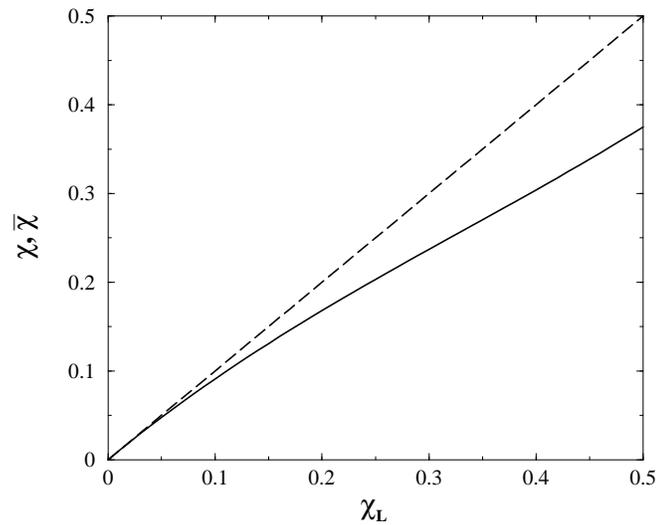}
    \caption{Nonlinear dependence of the susceptibility $\bar\chi=\chi_L(1-\chi_L+\chi_L^2)$
    (solid line) on the Langevin susceptibility $\chi_L$. The dashed line shows the
    linear function $\chi=\chi_L$.}
    \label{fig:1}
  \end{center}
\end{figure}

The discussion about the range of validity of ${\bf f}_{{\rm K}}$ in \cite{liu_comment00}
is based on the {\it simultaneous} approximation that $\chi\sim\rho$ and $\bar\chi\sim\rho$.
The concurrent correctness of both relations has to be checked very cautiously.
Since $\rho=\chi_L/\alpha$ with constant $\alpha$, the proportionality to the density
$\rho$ is equivalent with the proportionality to the Langevin susceptibility
$\chi_L$.  Restricting the dependence of the susceptibilities on $\chi_L$ up to the
third order, one has
\begin{equation}
\label{eq:6}
\chi = \chi_L\left[ 1+\beta_1\chi_L+\beta_2\chi_L^2\right]+O(\chi_L^4)
\end{equation}
and 
\begin{eqnarray}
\nonumber
\bar\chi &\simeq&\chi(1-\chi +\chi^2 -\cdots)\\
\label{eq:7}
&=& \chi_L\left[ 1+(\beta_1-1)\chi_L +(\beta_2+1-2\beta_1)\chi_L^2\right]+O(\chi_L^4)\; ,
\end{eqnarray}
where the expansion~(\ref{eq:7}) is valid for $\chi\ll 1$ only. Assuming a
linear dependence of $\chi$ on $\chi_L$, i.e. $\beta_1= \beta_2=0$,
the expansion~(\ref{eq:7}) implies necessarily that $\bar\chi$ depends on higher order
terms of $\chi_L$. Figure~\ref{fig:1}
shows the {\it linear} behaviour of $\chi=\chi_L$ (dashed line) and the {\it nonlinear}
behaviour of $\bar\chi=\chi_L(1-\chi_L+\chi_L^2)$ (solid line) for
$0\leq \chi_L\leq 0.5$. A nonlinear dependence of $\bar\chi$ on $\chi_L$ in
the region, where $\chi\sim\chi_L$ holds, is confirmed also by measurements (see Fig.~1(b)
in \cite{luo_comment00}).

From Fig.~\ref{fig:1} it becomes evident that in the region $\chi=\chi_L$ a subregion
$\chi_L\leq 0.06$ exists, where additionally $\bar\chi=\chi_L$ is fulfilled.
Inserting $\chi=\chi_L$ in Eqs.~(\ref{eq:2},\ref{eq:5}), the resulting force
density
\begin{equation}
\label{eq:8}
{\bf f}_{{\rm K}}(\chi)={\bf f}_{{\rm V}}(\chi)=-{B_{ext}^2\over \mu_0}
  {\chi_L\grad\chi_L\over (1+\chi_L)^3}
\end{equation}
is nonzero. This agreement confirms the above general statement that the usual
and the variant form of the Kelvin force density are equivalent provided they are
functions of $\chi$ $\bigr($see Eqs.~(\ref{eq:2},\ref{eq:5})$\bigr)$.
But inserting $\chi=\chi_L$, ${\bf M}=\chi{\bf H}_{int}$ in Eq.~(\ref{eq:1})
and $\bar\chi=\chi_L$, ${\bf M}=(\bar\chi/\mu_0){\bf B}_{int}$ in Eq.~(\ref{eq:4}),
respectively, one gets
\begin{equation}
\label{eq:9}
{\bf f}_{{\rm K}} =\mu_0({\bf M}\grad){\bf H}_{int}
\end{equation}
versus
\begin{equation}
\label{eq:10}
{\bf f}_{{\rm V}} =({\bf M}\grad){\bf B}_{int}\; .
\end{equation}
Where the first expression gives a nonzero force density equal to~(\ref{eq:8}), it is
zero in the second case. The reason for the difference between the nonzero result of
Eq.~(\ref{eq:8}) and the zero one of Eq.~(\ref{eq:10}) is the following:
the variant form of the Helmholtz force ~(\ref{eq:4}) is a direct function of any
approximation of $\bar\chi$ whereas the correct definition of $\bar\chi$ was
incorporated into ${\bf f}_{{\rm V}}(\chi)$ $\bigr($see Eq.~(\ref{eq:5})$\bigr)$.
That is the deeper reason why approximations cause differences if the two formulae for
the Helmholtz force are used.
It has to be noted that this discrepancy is limited to a
small subregion $\chi=\chi_L\leq 0.06$ which is outside the usual experimental
fluids. The lowest susceptibility of commercially available fluids is 0.13
\cite{ferrofluidics}.

For $\chi=\chi_L > 0.06$ a truncation in the expansion of $\bar\chi$ after the linear order
is deficient (see Fig.~\ref{fig:1}). If one inserts instead
the entire term $\bar\chi=\chi_L(1-\chi_L+\chi_L^2)$ in Eq.~(\ref{eq:4}), one obtains 
a nonzero force density also for ${\bf f}_{{\rm V}}$,
\begin{equation}
\label{eq:11}
{\bf f}_{{\rm V}} =-{B_{ext}^2\over \mu_0}(1-3\chi_L)\chi_L\grad\chi_L\; ,
\end{equation}
which is a good approximation of~(\ref{eq:8}) for small $\chi_L$.

These theoretical calculations as well as the experimental measurements in
\cite{luo_comment00}
show apparently that (i) a linear dependence of $\chi$ on the density results not
necessarily in a linear dependence of $\bar\chi$ on the density and (ii) nonlinear
contributions of $\rho$ are relevant for $\bar\chi$ even in the region $\chi\ll 1$.
(iii) The two formulae for the Helmholtz force are  entirely equivalent. It
is only when one introduces approximations that differences arise in a small subregion,
$\chi=\bar\chi=\chi_L\leq 0.06$.

\end{document}